\documentclass{article}

\usepackage[english]{babel}

\usepackage[letterpaper,top=2cm,bottom=2cm,left=3cm,right=3cm,marginparwidth=1.75cm]{geometry}

\usepackage{amsmath}
\usepackage{graphicx}
\usepackage[colorlinks=true, allcolors=blue]{hyperref}

\title{First thought on a high-intensity $K_S$ experiment}
\author{Radoslav Marchevski, \newline \textit{Weizmann Institute of Science}, Rehovot, Israel}

\begin{document}
\maketitle

\begin{abstract}
The $K \rightarrow \mu\mu$ decays have recently been identified as another golden kaon physics mode alongside the rare $K \rightarrow\pi\nu\bar{\nu}$ processes. These golden modes provide precision tests of the Standard Model with very high sensitivity to New Physics. The presented study is exploring the possibility to address the $K_L - K_S \rightarrow \mu^+\mu^-$ interference experimentally and outlines the challenges associated with such an ambitious project for the far future. A next-generation experiment at the intensity frontier is required that should be capable of collecting a large sample of $\mathcal{O}(10^{14} - 10^{15})$ $K_L$ and $K_S$ decays. Challenges related to the beamline design and detector technology need to be overcome if we want to address this mode experimentally. A significant background suppression of $K_S \rightarrow \pi^+\pi^-$ and radiative $K_L \rightarrow \mu^+\mu^-\gamma$ decays is imperative for a few $\%$ measurement, which would require excellent kinematic resolution and efficient photon detection. The first attempt at a possible experimental setup to measure this effect is presented. Last but not least, a huge number of neutral particles produced offers the possibility to study a plethora of other rare $K_L$, $K_S$ decays as well as hyperon decays enhancing the physics motivation for such an initiative.
\end{abstract}

\section{Introduction}
Over the past few years, kaon physics is attracting more and more attention. Recent experimental results on kaon Flavour Changing Neutral Currents (FCNCs) from NA62~\cite{NA62:pnn_1,NA62:pnn_2,NA62:pnn_3} and LHCb~\cite{LHCb:ksmumu_run2} at CERN, and KOTO~\cite{PhysRevLett.122.021802, PhysRevLett.126.121801} in Japan have produced a wide range of important results, which triggered a broad theoretical interest. Recently it has been shown that a measurement of the interference between $K_L \rightarrow \mu^+\mu^-$ and $K_S \rightarrow \mu^+\mu^-$ decays can be used to extract the $CP$ violation parameter $\eta$ with a theoretical precision of 1$\%$\cite{PhysRevLett.119.201802, Dery:2021mct}. The exclusive sensitivity of the interference to the short-distance $CP$-violating part of the $K \rightarrow \mu^+\mu^-$ amplitude turns the $K \rightarrow \mu^+\mu^-$ process into yet another golden rare kaon channel. This process can precisely probe the CKM structure of the SM, and at the same time offers a large sensitivity to contributions from physics beyond the Standard Model. A measurement of the $K_S - K_L$ interference effect ($BR_{eff} \sim 8 \times 10^{-10}$) to a few $\%$ precision to match the theoretical knowledge of this process is an essential flavour physics objective, and effort must be spent to address it. Such a measurement will be an extremely challenging task far beyond the reach of modern kaon physics experiments. It will only be possible at next-generation kaon experiments at the intensity frontier.
\section{High-intensity neutral kaon experiment}

\begin{figure}[ht]
\centering
\includegraphics[width=0.95\textwidth]{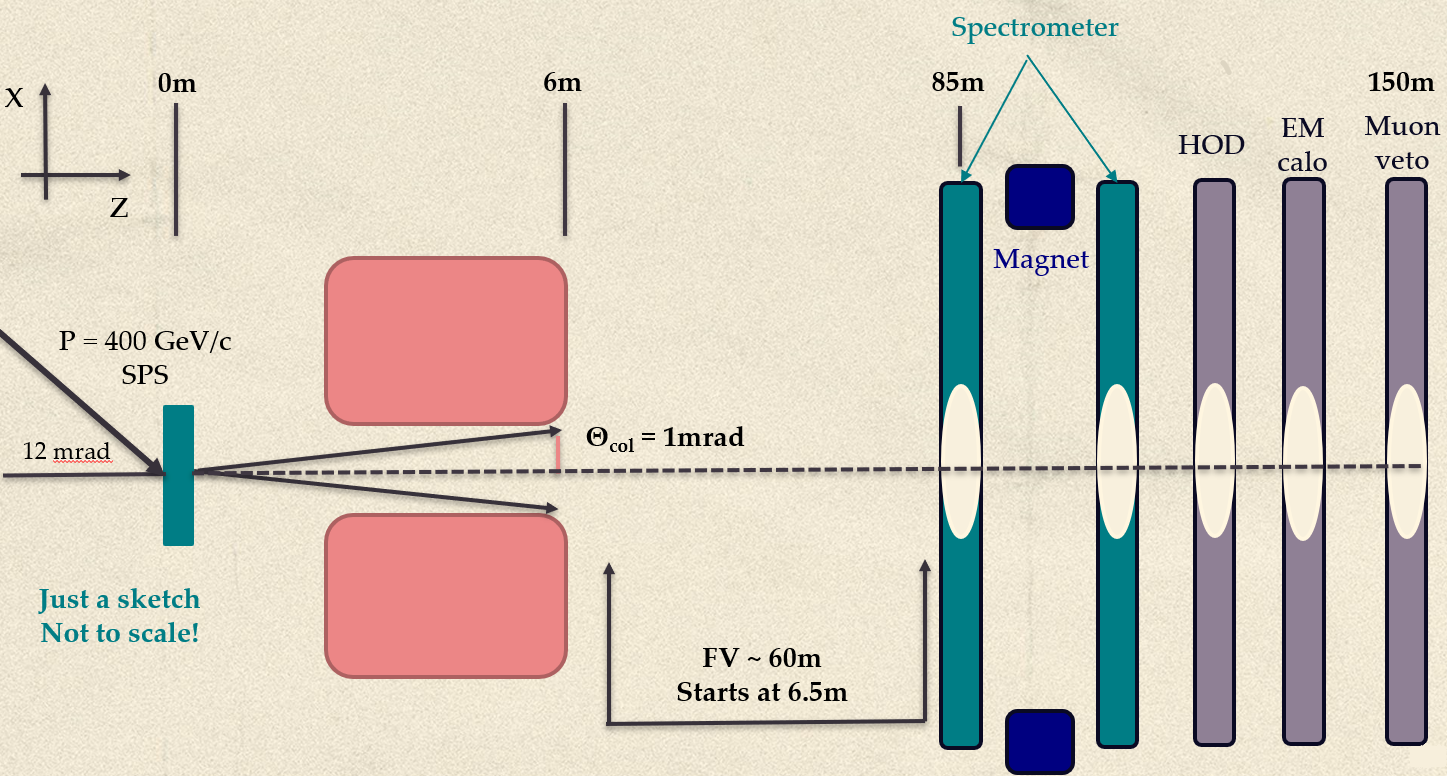}
\caption{\label{fig:exp}Sketch of a possible experimental layout for the high-intensity neutral kaon experiment.}
\end{figure}
The first attempt at an experimental setup to measure extremely rare neutral kaon decays is inspired by the design of the successful NA62 experiment\cite{NA62:2017rwk}. The setup uses a neutral secondary beam instead of a charged one. The experiment will use a 400~GeV$/c$ primary proton beam extracted from the CERN SPS accelerator impinging on a Beryllium target at a 12~mrad incident angle. The secondary beam opening angle is 1~mrad with respect to the center of the target defined by a 6~m long collimating system. The large incident angle of the primary beam will result in a soft kaon momentum spectrum, resulting in a geometrical acceptance of 30-40$\%$ for two-body kaon decays. The collimator is followed by a 60~m long decay region and a spectrometer, both located in the same vacuum tank. A charged hodoscope will allow precise timing measurements of the traversing charged particles. A calorimetric system complemented by a fast muon veto detector will provide particle identification capability to separate muons from pions and electrons. A sketch of the experimental layout is presented in Figure~\ref{fig:exp}.

An option to increase signal acceptance and provide parallel tracks for two-body decays by adding a second magnet to the experimental setup can also be studied. This double-bend technique was successfully employed by the E871 experiment at the Brookhaven National Laboratory~\cite{PhysRevLett.81.4309, PhysRevLett.81.5734} to reduce background from semileptonic kaon decays and can be an important improvement of the proposed experimental setup.
\section{Sensitivity and charged particle rates}

A toy simulation is developed to estimate the sensitivity of the setup. The differential momentum spectrum of the secondary kaon beam is generated using the Malensek parametrization~\cite{vanDijk:2018aaa}. The spectrum depends on the incident angle of the primary beam and the solid angle covered by the collimator opening. The signal process is generated according to the time-dependent rate~\cite{Dery:2021mct} and include three components: $K_S\rightarrow \mu^+\mu^-$; $K_L \rightarrow \mu^+\mu^-$; $K_S - K_L \rightarrow \mu^+\mu^-$ interference. After the decay, the resulting muons are further propagated through the whole experimental setup to estimate the signal acceptance. The main kinematic quantity used is the di-muon invariant masts, $M_{\mu\mu} = \sqrt{(P_{\mu 1} + P_{\mu 2})^2}$, where $P_{\mu 1}$ and $P_{\mu 2}$ are the momenta of the two muon tracks. The $M_{\mu\mu}$ distribution for the signal is a peak centered at the neutral kaon mass. In the simulation, the momentum and angular resolution of the spectrometer for muon tracks are assumed to be the same as for the existing straw tube tracker of NA62. A smearing factor is applied, resulting in an invariant mass resolution of $\sigma_{M} \sim 1.9~\text{MeV}/c^2$ for di-muon events. The signal region is defined in the 492--504~MeV$/c^2$ range. The geometrical signal selection results in 40$\%$ acceptance for the signal, which is quite encouraging. Further corrections are applied to account for additional background-suppressing conditions, projected DAQ, Trigger, and detector efficiencies bringing signal efficiency down to 15$\%$. This number should represent a more realistic estimate of the true signal efficiency but it relies on a lot of assumptions and results in significant uncertainties. The signal yield of the experimental setup described above is between 75 and 300 interference events/year after applying the $15\%$ signal efficiency. The expected number of interference events in the final experiment can`t reliably be computed because it depends heavily on the choice of a beam setup (incident angle, collimation scheme) and the strong phase, $\varphi_0$, governing the size of the interference effect. Optimizations of the beamline are essential to determine the ultimate sensitivity of the experiment.

The main background is produced by $K_S\rightarrow \pi^+\pi^- (BR\approx 70\%)$ decays either through a double $\pi \rightarrow \mu $ misidentification or two consecutive $\pi^+ \rightarrow \mu^+\nu_\mu$ decays. Both backgrounds must be suppressed by at least $10^{11}$, which will present significant experimental challenges. These challenges can be addressed by: strong particle identification, and excellent kinematic resolution. Another source of background will be $K_L \rightarrow \mu^+\mu^-\gamma(BR\approx3.6\times 10^{-7})$ decay, where the presence of an additional photon and the small branching ratio should be sufficient to bring the background to the desirable level.

\begin{figure}[ht]
\centering
\includegraphics[width=0.95\textwidth]{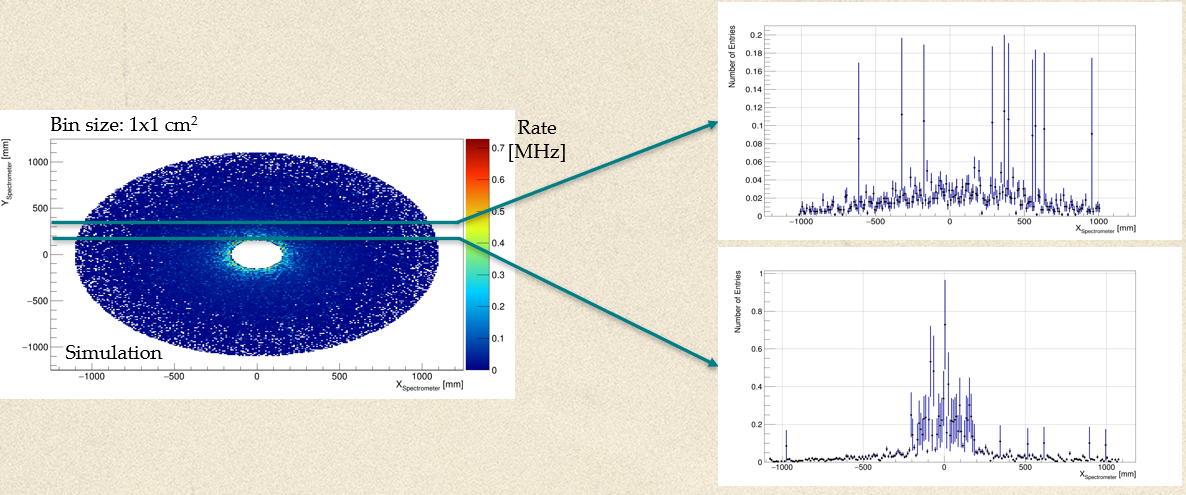}
\caption{\label{fig:rates}Sketch of a possible experimental layout for the high-intensity neutral kaon experiment.}
\end{figure}

%Improve the two paragraphs below
A main challenge for the experiment will be the particle rate in the detectors, which is primarily generated by $K_S$ and $\Lambda$ decays. Assuming $10^{19}$ POT the rate of charged particles at the first spectrometer station, located at a $z =$ 85~m from the primary target is about 1~GHz over a surface area of 3.7~m$^2$. The rate is heavily dependent on the distance from the beam (see Figure~\ref{fig:rates}). The high rate in the central parts of the detector is produced primarily by $\Lambda \rightarrow p\pi^-$ decays, where the proton takes a larger part of the momentum and primarily traverses the central part of the detector. The highly non-uniform rate imposes technological challenges on the required detectors. While the charged particle rates are low at the outer parts of the detector (50-100~KHz$/\text{cm}^2$) in the central parts the rates can reach up to 0.7-1 MHz$/\text{cm}^2$, an order of magnitude higher. This challenge can be solved by developing high-granularity detectors with different technology as a function of radius. However, the interface between the different detector materials will not be trivial. Solid-state detectors might be the solution and we can look for solutions required for detectors at the HL-LHC which impose similar requirements. Finally, the large rates also require a novel readout system. A standard TDAQ chain involving a low-level hardware trigger followed by an online software-based system might not be feasible. Solutions adopting a purely software-trigger system will be investigated.
%\begin{table}
%\centering
%\begin{tabular}{l|r}
%Item & Quantity \\\hline
%Widgets & 42 \\
%Gadgets & 13
%\end{tabular}
%\caption{\label{tab:widgets}An example table.}
%\end{table}

\section{Areas for future development}
The presented sensitivity projections albeit quite crude are encouraging and warrant more serious feasibility studies. The number of expected interference events depends on the strong phase $\varphi_0$ and values that produce constructive interference are favourable. Work on the SM estimate gives a value of $\cos{\varphi_0} = 0.978 \pm 0.009$\cite{Avital_kmumu_kaon2022} for the strong phase, indicating maximal constructive interference and better signal sensitivity.

A more realistic design of the beamline should be developed. Optimizing the incident angle of the primary beam onto the target, collimation, and muon shielding will determine if the necessary statistics of $\mathcal{O}(10^3)$ interference events can be collected. To collect such a large number of rare events implies large rates in the detectors downstream of the fiducial volume. A dedicated R$\&$D program is required to provide tracking and calorimetry at the GHz regime. High-granularity detectors with a time resolution of about 100~ps or better are needed, which presents a technological challenge that must be tackled over the next years. The rate in the detectors will be highly non-uniform over the detection surface (see Figure~\ref{fig:rates}). The large particle rates close to the beam pipe would require the development of hybrid detectors using detection techniques with different rate capabilities as a function of the distance from the beam axis.

The detectors must also provide excellent spatial, momentum, and energy resolution to ensure the necessary suppression of background decays and accidentals. The sensitivity of the measurement can be heavily impacted by background contamination adding a systematic uncertainty to the extraction of the $\eta$. The background from $K_S \rightarrow \pi^+\pi^-$ and $K_L \rightarrow \mu^+\mu^-\gamma$ processes produce the main background from kaon decays. The experimental design will be heavily geared towards precise kinematic reconstruction, fast timing, and hermetic photon detection that must provide the necessary background rejection. The large rate of muons generated from the beamline can lead to accidental di-muon pairs and can be a large source of background. This background has not been addressed so far and must be one of the main concerns during the beamline optimization process.

The large amount of $\mathcal{O}(10^{14})$ neutral kaon decays and $\mathcal{O}(10^{13})$ $\Lambda$ baryon decays offers a great opportunity to discover and measure other very rare processes. Notable examples are the $K_L \rightarrow \pi^0l^+l^-$ and $K_L \rightarrow \mu e$ decays, which are of great theoretical interest because they can provide important constraints on various NP scenarios. Sensitivity studies for a broad range of processes must be performed and new ideas about interesting observables are welcome.

\section{Conclusions}
The study of $K \rightarrow \mu^+\mu^-$ decays presents a golden opportunity to obtain a clean determination of the $CP$-violating parameter $\eta$ from kaon physics, in addition to the golden $K_L \rightarrow \pi^0 \nu \bar{\nu}$ mode. The capabilities of the CERN facilities to deliver high-intensity kaon beams offer interesting prospects to measure the $K_S - K_L \rightarrow \mu^+\mu^-$ interference with a few $\%$ precision in the future. If such an experiment can be constructed it will enable a very broad physics program complementary to the $K \rightarrow \mu^+\mu^-$ studies. The large number of $3-4 \times10^{13}$ kaon decays/year will allow significant improvement in the precision of a wide range of kaon observables, as well as very sensitive searches addressing a broad range of NP scenarios. The large particle rates in the detectors impose severe technical challenges, which might need at least a decade for its solution. Very fast and highly granular detectors will be imperative and require a dedicated R$\&$D program. Innovative solutions can be found by exploring synergies with detector developments for experiments at the high-luminosity phase of the LHC, which will be essential for the next-generation kaon experiments at the intensity frontier.

\bibliographystyle{JHEP}
\bibliography{bibliography}

\end{document}